\documentstyle[12pt,epsfig]{article}

\catcode`\@=11
\def\@citex[#1]#2{%
\if@filesw \immediate \write \@auxout {\string \citation {#2}}\fi
\@tempcntb\m@ne \let\@h@ld\relax \def\@citea{}%
\@cite{%
  \@for \@citeb:=#2\do {%
    \@ifundefined {b@\@citeb}%
      {\@h@ld\@citea\@tempcntb\m@ne{\bf ?}%
      \@warning {Citation `\@citeb ' on page \thepage \space
undefined}}%
      {\@tempcnta\@tempcntb \advance\@tempcnta\@ne%
      \@tempcntb\number\csname b@\@citeb \endcsname \relax%
      \ifnum\@tempcnta=\@tempcntb 
        \ifx\@h@ld\relax%
          \edef \@h@ld{\@citea\csname b@\@citeb\endcsname}%
        \else%
          \edef\@h@ld{\ifmmode{-}\else--\fi\csname
b@\@citeb\endcsname}%
        \fi%
      \else
        \@h@ld\@citea\csname b@\@citeb \endcsname%
        \let\@h@ld\relax%
      \fi}%
    \def\@citea{,\penalty\@highpenalty\,}%
  }\@h@ld
}{#1}}

\def\@citeb#1#2{{[#1]\if@tempswa , #2\fi}}
\def\@citeu#1#2{{$^{#1}$\if@tempswa , #2\fi }}
\def\@citep#1#2{{#1\if@tempswa , #2\fi}}

\def\bcites{         
        \catcode`\@=11
        \let\@cite=\@citeb
        \catcode`\@=12
}

\def\upcites{         
        \catcode`\@=11
        \let\@cite=\@citeu
        \catcode`\@=12
}

\def\plaincites{      
        \catcode`\@=11
        \let\@cite=\@citep
        \catcode`\@=12
}

\newcount\hour
\newcount\minute
\newtoks\amorpm
\hour=\time\divide\hour by 60
\minute=\time{\multiply\hour by 60 \global\advance\minute by-\hour}
\edef\standardtime{{\ifnum\hour<12 \global\amorpm={am}%
        \else\global\amorpm={pm}\advance\hour by-12 \fi
        \ifnum\hour=0 \hour=12 \fi
        \number\hour:\ifnum\minute<10
0\fi\number\minute\the\amorpm}}
\edef\militarytime{\number\hour:\ifnum\minute<10
0\fi\number\minute}

\def\draftlabel#1{{\@bsphack\if@filesw {\let\thepage\relax
   \xdef\@gtempa{\write\@auxout{\string
      \newlabel{#1}{{\@currentlabel}{\thepage}}}}}\@gtempa
   \if@nobreak \ifvmode\nobreak\fi\fi\fi\@esphack}
        \gdef\@eqnlabel{#1}}
\def\@eqnlabel{}
\def\@vacuum{}
\def\marginnote#1{}
\def\draftmarginnote#1{\marginpar{\raggedright\scriptsize\tt#1}}
\def\draft{
        \pagestyle{plain}
        \overfullrule=2pt
        \oddsidemargin -.5truein
        \def\@oddhead{\sl \phantom{\today\quad\militarytime} \hfil
        \smash{\Large\sl DRAFT} \hfil \today\quad\militarytime}
        \let\@evenhead\@oddhead
        \let\label=\draftlabel
        \let\marginnote=\draftmarginnote
        \def\ps@empty{\let\@mkboth\@gobbletwo
        \def\@oddfoot{\hfil \smash{\Large\sl DRAFT} \hfil}
        \let\@evenfoot\@oddhead}

\def\@eqnnum{(\theequation)\rlap{\kern\marginparsep\tt\@eqnlabel}%
        \global\let\@eqnlabel\@vacuum}  }

\def\blackfonts{
        \font\blackboard=msbm10 scaled\magstep1
        \font\blackboards=msbm8
        \font\blackboardss=msbm6
}

\def\nblack{            
        \def\ZZ{{Z \n{10} Z}}
        \def\NN{{N \n{14} N}}
        \def\CC{{C \n{11} C}}
        \def\RR{{R \n{11} R}}
        \def\QQ{{Q \n{12} Q}}
        \def\PP{{P \n{11} P}}
}

\def\prep{         
        \catcode`\@=11
        \input art10.sty
        \catcode`\@=12
        
        \let\small\null
        \def\blackfonts{
                \font\blackboard=msbm10
                \font\blackboards=msbm7
                \font\blackboardss=msbm5
        }
        \let\sl\it
        \twocolumn
        \sloppy
        \voffset=-2.54truecm
        \hoffset=-2.54truecm
        \flushbottom
        \parindent 1em
        \leftmargini 2em
        \leftmarginv .5em
        \leftmarginvi .5em
        \marginparwidth 48pt
        \marginparsep 10pt
        \setlength{\columnsep}{2truecm}
        \setlength{\textwidth}{25.4truecm}
        \setlength{\textheight}{17truecm}
>       \baselineskip=16pt
        \oddsidemargin .18truein
        \evensidemargin .17truein
}

\def\eqalign#1{\null\,\vcenter{\openup\jot\m@th
  \ialign{\strut\hfil$\displaystyle{##}$&$\displaystyle{{}##}$\hfil
      \crcr#1\crcr}}\,}
\def\eqalignno#1{\displ@y \tabskip\centering
  \halign
to\displaywidth{\hfil$\@lign\displaystyle{##}$\tabskip\z@skip
    &$\@lign\displaystyle{{}##}$\hfil\tabskip\centering
    &\llap{$\@lign##$}\tabskip\z@skip\crcr
    #1\crcr}}

\def\section{\@startsection {section}{1}{\z@}{3.ex plus 1ex minus
 .2ex}{2.ex plus .2ex}{\large\bf}}
\def\subsection{\@startsection{subsection}{2}{\z@}{2.75ex plus 1ex
minus
 .2ex}{1.5ex plus .2ex}{\bf}}

\def\appendix{{\newpage\section*{Appendix}}\let\appendix\section%
        {\setcounter{section}{0}
        \gdef\thesection{\Alph{section}}}\section}

\def\abstract{\if@twocolumn
\section*{Abstract}
\else 
\begin{center}
{\bf Abstract\vspace{-.5em}\vspace{0pt}}
\end{center}
\quotation
\fi}

\catcode`\@=12

\def\ibid{{\it ibid.\/}}

\newcommand{\beq}{\begin{equation}}
\newcommand{\eeq}{\end{equation}}
\newcommand{\be}{\beq}
\newcommand{\ee}{\eeq}
\newcommand{\beqa}{\begin{eqnarray}}
\newcommand{\eeqa}{\end{eqnarray}}
\newcommand{\bea}{\beqa}
\newcommand{\eea}{\eeqa}

\def\noj#1,#2,{{\bf #1} (19#2)\ }
\def\jou#1,#2,#3,{{\sl #1\/ }{\bf #2} (19#3)\ }
\def\ann#1,#2,{{\sl Ann.\ Physics\/ }{\bf #1} (19#2)\ }
\def\cmp#1,#2,{{\sl Comm.\ Math.\ Phys.\/ }{\bf #1} (19#2)\ }
\def\ma#1,#2,{{\sl Math.\ Ann.\/ }{\bf #1} (19#2)\ }
\def\jd#1,#2,{{\sl J.\ Diff.\ Geom.\/ }{\bf #1} (19#2)\ }
\def\invm#1,#2,{{\sl Invent.\ Math.\/ }{\bf #1} (19#2)\ }
\def\cq#1,#2,{{\sl Class.\ Quantum Grav.\/ }{\bf #1} (19#2)\ }
\def\cqg#1,#2,{{\sl Class.\ Quantum Grav.\/ }{\bf #1} (19#2)\ }
\def\ijmp#1,#2,{{\sl Int.\ J.\ Mod.\ Phys.\/ }{\bf A#1} (19#2)\ }
\def\jmphy#1,#2,{{\sl J.\ Geom.\ Phys.\/ }{\bf #1} (19#2)\ }
\def\jams#1,#2,{{\sl J.\ Amer.\ Math.\ Soc.\/ }{\bf #1} (19#2)\ }
\def\grg#1,#2,{{\sl Gen.\ Rel.\ Grav.\/ }{\bf #1} (19#2)\ }
\def\mpl#1,#2,{{\sl Mod.\ Phys.\ Lett.\/ }{\bf A#1} (19#2)\ }
\def\nc#1,#2,{{\sl Nuovo Cim.\/ }{\bf #1} (19#2)\ }
\def\np#1,#2,{{\sl Nucl.\ Phys.\/ }{\bf B#1} (19#2)\ }
\def\pl#1,#2,{{\sl Phys.\ Lett.\/ }{\bf #1B} (19#2)\ }
\def\pla#1,#2,{{\sl Phys.\ Lett.\/ }{\bf #1A} (19#2)\ }
\def\pr#1,#2,{{\sl Phys.\ Rev.\/ }{\bf #1} (19#2)\ }
\def\prd#1,#2,{{\sl Phys.\ Rev.\/ }{\bf D#1} (19#2)\ }
\def\prl#1,#2,{{\sl Phys.\ Rev.\ Lett.\/ }{\bf #1} (19#2)\ }
\def\prp#1,#2,{{\sl Phys.\ Rept.\/ }{\bf #1C} (19#2)\ }
\def\ptp#1,#2,{{\sl Prog.\ Theor.\ Phys.\/ }{\bf #1} (19#2)\ }
\def\ptpsup#1,#2,{{\sl Prog.\ Theor.\ Phys.\/ Suppl.\/ }{\bf #1}
(19#2)\ }
\def\rmp#1,#2,{{\sl Rev.\ Mod.\ Phys.\/ }{\bf #1} (19#2)\ }
\def\yadfiz#1,#2,#3[#4,#5]{{\sl Yad.\ Fiz.\/ }{\bf #1} (19#2) #3%
\ [{\sl Sov.\ J.\ Nucl.\ Phys.\/ }{\bf #4} (19#2) #5]}
\def\zh#1,#2,#3[#4,#5]{{\sl Zh..\ Exp.\ Theor.\ Fiz.\/ }{\bf #1}
(19#2) #3%
\ [{\sl Sov.\ Phys.\ JETP\/ }{\bf #4} (19#2) #5]}

\def\beq{\begin{equation}}
\def\eeq{\end{equation}}
\def\beqar{\begin{eqnarray}}
\def\eeqar{\end{eqnarray}}

\def\nfrac#1#2{{\displaystyle{\vphantom1\smash{\lower.5ex\hbox{\sma
ll$#1$}}%
        \over\vphantom1\smash{\raise.25ex\hbox{\small$#2$}}}}}

\def\n#1{\mskip-#1mu}

\def\lae{\mathrel{\mathop{\smash{\lower .5 ex
\hbox{$\stackrel<\sim$}}}}}
\def\lae{\mathrel{\mathop{\smash{\lower .5 ex
\hbox{$\stackrel>\sim$}}}}}

\def\l:{\mathopen{:}\,}
\def\r:{\,\mathclose{:}}

\catcode`\@=11
\def\theequation{\arabic{equation}}
\catcode`\@=12

\nblack
\bcites

\nblack

\catcode`\@=11
\def\theequation{\thesection.\arabic{equation}}
\@addtoreset{equation}{section}
\@addtoreset{footnote}{section}
\@addtoreset{footnote}{subsection}
\catcode`\@=12

\newcommand{\beqn}{\begin{equation}}
\newcommand{\eeqn}{\end{equation}}
\newcommand{\beqnarray}{\begin{eqnarray}}
\newcommand{\eeqnarray}{\end{eqnarray}}
\newcommand {\bear} [1] {\begin {array} {#1}}
\newcommand {\ear} {\end {array}}

\newcommand {\beqarn} {\begin{eqnarray*}}
\newcommand {\eeqarn} {\end{eqnarray*}}

\newcommand {\mrm} [1] {\mathrm {#1}}

\newcommand {\myref} [1]        %
        {%
        \begin{thebibliography} {99}    %
                        {#1}    %
        \end {thebibliography}}

\def\emph#1{{\em #1}}
\def\mathbf#1{{\bf #1}}
\def\mathrm#1{{\rm #1}}
\def\mathit#1{{\it #1}}

\catcode`\@=11
\@addtoreset{footnote}{section}
\catcode`\@=12

\newcounter     {exinsert}      [subsection]

\renewcommand {\theexinsert}    %
{       \thesubsection.\arabic{equation}}

\renewcommand {\theequation} {\thesection.\arabic{equation}}

\catcode`\@=11
\@addtoreset{equation}{section}
\catcode`\@=12

\newenvironment {exinsert} [1]  %
{       
        \begin {quotation}      %
        \refstepcounter {equation}      %
        {\bf {#1}} \theequation. \,\,   %
}
{       
        \end {quotation}
}

\newcommand {\bprop}
{\begin {exinsert} {Proposition}
}
\newcommand {\eprop} {\end {exinsert} }

\newcommand {\bexe} {\begin {exinsert} {Exercise}}
\newcommand {\eexe} {\end {exinsert} }

\newcommand {\bexa} {\begin {exinsert} {Example}}
\newcommand {\eexa} {\end {exinsert} }

\newcommand {\literal} [1] {{ \mbox {$\mrm {#1}$}}}

\newcommand {\grad} {\nabla}

\newcommand {\ncmd} [2] {\newcommand {#1} {#2}}

\ncmd {\vgrad} {{\vec \grad}}

\ncmd {\Mv} {{\cal M}_V}
\ncmd {\Mh} {{\cal M}_H}

\ncmd {\adj} {\literal {adj}}   
\ncmd {\Mp} {M_\literal {planck}}       

\ncmd {\x} {$\times$}

\newcommand {\btab} [3]
  {\begin{table} [htb]
   \begin{center}
   \caption {#2}        \label {tab:#1}
   \begin {tabular} {#3}
   \hline}

\newcommand {\etab}
        {  \hline   \end {tabular}
   \end{center}
  \end{table}}

\ncmd {\bR} {\textbf {R}}
\ncmd {\bC} {\textbf {C}}

\ncmd {\CQ} {\cal Q}

\def\jou#1,#2,#3,{{\sl #1\/ }{\bf #2} (19#3)\ }
\def\ibid#1,#2,{{\sl ibid.\/ }{\bf #1} (19#2)\ }
\def\am#1,#2,{{\sl Acta. Math.\/ } {\bf #1} (19#2)\ }
\def\annp#1,#2,{{\sl Ann.\ Physics\/ }{\bf #1} (19#2)\ }
\def\cmp#1,#2,{{\sl Comm.\ Math.\ Phys.\/ }{\bf #1} (19#2)\ }
\def\cqg#1,#2,{{\sl Class.\ Quantum Grav.\/ }{\bf #1} (19#2)\ }
\def\dm#1,#2,{{\sl Duke\ Math.\ J.\/ }{\bf #1} (19#2)\ }
\def\ijmpa#1,#2,{{\sl Int.\ J.\ Mod.\ Phys.\/ }{\bf A#1} (19#2)\ }
\def\invm#1,#2,{{\sl Invent.\ Math.\/ }{\bf #1} (19#2)\ }
\def\jhep#1,#2,{{\sl JHEP\/ }{\bf #1} (19#2)\ }
\def\jmp#1,#2,{{\sl J.\ Geom.\ Phys.\/ }{\bf #1} (19#2)\ }
\def\jams#1,#2,{{\sl J.\ Diff.\ Geom.\/ }{\bf #1} (19#2)\ }
\def\jdg#1,#2,{{\sl J.\ Amer.\ Math.\ Soc.\/ }{\bf #1} (19#2)\ }
\def\jpa#1,#2,{{\sl J.\ Phys.\ A.\/ }{\bf #1} (19#2)\ }
\def\grg#1,#2,{{\sl Gen.\ Rel.\ Grav.\/ }{\bf #1} (19#2)\ }
\def\mpla#1,#2,{{\sl Mod.\ Phys.\ Lett.\/ }{\bf A#1} (19#2)\ }
\def\ma#1,#2,{{\sl Math.\ Ann.\/ } {\bf #1} (19#2)\ }
\def\nc#1,#2,{{\sl Nuovo\ Cim.\/ }{\bf #1} (19#2)\ }
\def\npb#1,#2,{{\sl Nucl.\ Phys.\/ }{\bf B#1} (19#2)\ }
\def\plb#1,#2,{{\sl Phys.\ Lett.\/ }{\bf #1B} (19#2)\ }
\def\pla#1,#2,{{\sl Phys.\ Lett.\/ }{\bf #1A} (19#2)\ }
\def\pnas#1,#2,{{\sl Proc..Nat.Acad.Sci.\/ }{\bf #1} (19#2)\ }
\def\prev#1,#2,{{\sl Phys.\ Rev.\/ }{\bf #1} (19#2)\ }
\def\prd#1,#2,{{\sl Phys.\ Rev.\/ }{\bf D#1} (19#2)\ }
\def\prl#1,#2,{{\sl Phys.\ Rev.\ Lett.\/ }{\bf #1} (19#2)\ }
\def\prpt#1,#2,{{\sl Phys.\ Rept.\/ }{\bf #1C} (19#2)\ }
\def\ptp#1,#2,{{\sl Prog.\ Theor.\ Phys.\/ }{\bf #1} (19#2)\ }
\def\ptpsup#1,#2,{{\sl Prog.\ Theor.\ Phys.\/ Suppl.\/ }{\bf #1} (19#2)\ }
\def\rmp#1,#2,{{\sl Rev.\ Mod.\ Phys.\/ }{\bf #1} (19#2)\ }
\def\sm#1,#2,{{\sl Selec. Math.\/ }{\bf #1} (19#2)\ }
\def\yadfiz#1,#2,#3[#4,#5]{{\sl Yad.\ Fiz.\/ }{\bf #1} (19#2) #3%
\ [{\sl Sov.\ J.\ Nucl.\ Phys.\/ }{\bf #4} (19#2) #5]}
\def\zh#1,#2,#3[#4,#5]{{\sl Zh.\ Exp.\ Theor.\ Fiz.\/ }{\bf #1} (19#2) #3%
\ [{\sl Sov.\ Phys.\ JETP\/ }{\bf #4} (19#2) #5]}

\def\cft {{\it CFT}}
\def\ads {{\it AdS}}

\begin {document}


\begin{titlepage}

\begin{center}
\hfill
UCB-PTH-98/43,~~
LBNL-42229\\
\hfill hep-th/9812046

\vskip 1.5 cm
{\LARGE \bf String Theory on AdS$_3$}
\vskip 1 cm
{\large Jan de\thinspace Boer, Hirosi Ooguri, 
Harlan Robins and Jonathan Tannenhauser}\\
\vskip 1cm
{Department of Physics,
University of California at Berkeley,\\
Berkeley, California 94720}\\
\vskip .3cm
{and}
\vskip .3cm
{Theoretical Physics Group, Mail Stop 50A-5101,\\
Lawrence Berkeley National Laboratory,\\
Berkeley, California 94720}\\

\end{center}

\vskip 0.5 cm
\begin{abstract}
It was shown by Brown and Henneaux that the classical theory 
of gravity on $AdS_3$ has an infinite-dimensional symmetry group
forming a Virasoro algebra. More recently, Giveon, Kutasov and Seiberg
(GKS) constructed the corresponding Virasoro generators in the 
first-quantized string theory on $AdS_3$. 
In this paper, we explore various aspects of string theory on $AdS_3$ and
study the relation between these two works. We show how
semi-classical properties of the string theory reproduce many
features of the \ads/\cft\ duality. 
Furthermore, we examine how the Virasoro symmetry of Brown and
Henneaux is realized in string theory, 
and show how it leads to the Virasoro Ward identities 
of the boundary \cft. The Virasoro generators of GKS emerge
naturally in this analysis. 
Our work clarifies several aspects of the GKS construction: 
why the Brown-Henneaux Virasoro algebra can be realized on the
first-quantized Hilbert space, to what extent the free-field 
approximation is valid, and why the 
Virasoro generators act on the string worldsheet 
localized near the boundary of $AdS_3$. On the other hand, we
find that the way the central charge of the Virasoro algebra is
generated is different from the mechanism proposed by GKS.
\end{abstract}

\end{titlepage}

\section{Introduction}

It was shown by Brown and Henneaux \cite{bh} that
the semi-classical theory of gravitation on three-dimensional
anti-de Sitter space (\ads$_3$) possesses an infinite-dimensional 
symmetry algebra of Virasoro type. 
The realization of this Virasoro algebra has 
recently been clarified in light of the 
\ads/\cft\ duality \cite{mal,gkp,w}, 
according to which string/$M$-theory on
a $(p+1)$-dimensional anti-de Sitter space times a
compact space is equivalent to a $p$-dimensional conformal field
theory (\cft$_p$). The case of \ads$_3$ was studied in more detail in
\cite{str,ms,mart,deboer}. 
More recently, Giveon, Kutasov and Seiberg (GKS)
found that the Brown-Henneaux Virasoro algebra is 
realized on the first-quantized string theory
on \ads$_3$, shedding further light
on the duality \cite{gks}.

The main purpose of this paper is to clarify the relation between the 
Brown-Henneaux Virasoro algebra and the Virasoro generators constructed 
by GKS. In the construction of GKS, the Virasoro 
algebra acts on the first-quantized string Hilbert space. 
However, as shown in \cite{str,ms,mart,deboer}, 
the Brown-Henneaux Virasoro 
operators are creation and annihilation operators of gravitons 
in $AdS_3$, and as such they are realized on the second-quantized Hilbert 
space of strings. It was not clear how to reconcile these two points
of view. In addition, GKS assume that the string worldsheet 
is localized near the boundary of $AdS_3$ and 
winds around the boundary. It is not obvious why we need (and need only)  
consider such worldsheet configurations. 
In this paper, we will give answers to these questions, and along the
way, we will recover many features of the \ads/\cft\ duality directly
from the worldsheet theory of strings on $AdS_3$.

As by-products of this analysis, we gain new insights into the structure 
of two-dimensional sigma models with non-compact target spaces such
as $AdS_3$. In the case of $AdS_3$ with the Euclidean-signature metric, 
the sigma model 
is unitary and its Hilbert space is equipped with a positive-definite 
inner product. Since $AdS_3$ has an $SO(3,1) \simeq SL(2,C)$ isometry group, 
the Hilbert space should decompose into a direct sum of unitary 
representations 
of $SL(2,C)$. To our surprise, we find that the \ads/\cft\ duality
implies that vertex operators of the sigma model belong to non-unitary 
representations of $SL(2,C)$. This is not a contradiction, and
appears to be a generic phenomenon in non-compact sigma models. 
It is closely related to the absence of the state-operator correspondence
in the Liouville model \cite{seib}, where it is known 
that normalizable 
states make up the Hilbert space, while non-normalizable states 
correspond to operators. 
In the $AdS_3$ case, unitary representations of $SL(2,C)$ are realized by
normalizable functions on $AdS_3$, whereas non-normalizable functions
give non-unitary representations. Thus it is reasonable, by
analogy with the Liouville model, that the vertex operators of the
sigma model belong to non-unitary representations.

This paper is organized as follows. 

In section 2, we briefly summarize the duality
between string theory on $AdS_3$ and conformal field theory 
in two dimensions.

In section 3, we discuss various aspects of the worldsheet 
theory of strings on Euclidean $AdS_3$, including the $SL(2,C)$
symmetry and the vertex operator construction. We show
that the worldsheet vertex operators are closely related to
bulk-boundary Green's functions in target space.

In section 4, we perform a semi-classical analysis of correlation
functions of primary fields, and show that the 
vertex operators are subject to the wave function renormalization
expected from the \ads/\cft\ duality and from the holographic
identification of the regularizations \cite{iruv}.
The worldsheet stretches to the boundary of \ads$_3$ at the
insertion points of the vertex operators, and can be viewed
as a thickening of the target-space Feynman diagram involving
bulk-boundary and bulk-bulk Green's functions. 

In section 5, we define the Virasoro generators as the graviton
vertex operators corresponding to Brown-Henneaux diffeomorphisms,
and explain why these vertex operators do not decouple from the theory. 

In section 6, we derive the Virasoro Ward identity of the boundary 
\cft\ and show how the Virasoro generators defined by
GKS \cite{gks} arise. 

In section 7, we consider the correlation function of two
boundary stress-energy tensors and explain how the central charge
appears. A crucial step is to consider disconnected worldsheets,
{\it i.e.}, second-quantized string theory.

We end in section 8 with some conclusions.

\section{The \ads$_3$/\cft$_2$ Duality}

Following \cite{ms}, we start with type IIB string theory on 
${\bf R}^4 \times {\bf R}^2
 \times M^4$, where $M^4$ is a compact manifold ($M^4 = T^4$ 
or K$_3$), and consider $Q_1$ fundamental strings on ${\bf R}^2$
and $Q_5$ {\it NS} fivebranes 
on ${\bf R}^2 \times M^4$. In the near-horizon limit,
the target space geometry is $AdS_3 \times S^3 \times M^4$, with a non-zero
${\it NS}$-${\it NS}$ 2-form. The curvature radius $l_{AdS}$ of 
$AdS_3$ is equal to $\sqrt{Q_5} l_s$ where $l_s$ is the string scale.
The string coupling constant on $AdS_3$ is
\beq
     g_3^2 = \frac{1}{Q_1 \sqrt{Q_5}}.
\eeq
Thus we have the following hierarchy of scales:
\beq
     l_{AdS} = \sqrt{Q_5} l_s = 4 Q_1 Q_5 l_p,
\eeq
where $l_p = \frac{1}{4} g_3^2 l_s$ is the three-dimensional Planck length. 
When $Q_1 Q_5 \gg 1$, quantum gravity effects are weak. 
Moreover when $Q_5 \gg 1$, the $\alpha'$-expansion of the worldsheet
theory becomes reliable. 

According to the \ads/\cft\
duality, this system is dual
to some two-dimensional conformal field theory (\cft$_2$). 
In the original work of Brown and Henneaux \cite{bh},
the central charge $c$ of the \cft$_2$ is given in the low-energy gravity approximation by
\beq
  c = \frac{3l_{AdS}}{2 l_p} = 6 Q_1 Q_5.
\eeq
This is consistent with the $S$-dual of the brane-configuration,
which is the D$1$-D$5$ system, whose field theory limit is 
a \cft$_2$ with $c=6Q_1Q_5$ \cite{sv}.

\section{Worldsheet Description of Strings on AdS$_3$}

\subsection{Action and Symmetry}

In Euclidean $AdS_3$, the bosonic part of the worldsheet Lagrangian is
\beq
  S_E = \frac{Q_5}{2\pi} \int d^2z ( \partial \phi \bar{\partial} \phi
   + e^{2 \phi} \partial \bar{\gamma} \bar{\partial}\gamma).
\label{action}
\eeq
Here $(\phi, \gamma, \bar{\gamma})$ are the coordinates on $AdS_3$.
The coordinate $\phi$ is real, while $\gamma$ and 
$\bar{\gamma}$ are complex conjugates.  The boundary of $AdS_3$ is 
located at $\phi = \infty$. In this sub-section we will summarize
known facts about this action, based on the earlier works \cite{gks,gav,gk}.

First of all, it is instructive to compare (\ref{action}) 
with the corresponding
action $S_L$ for $AdS_3$ with Lorentzian signature. Lorentzian-signature
$AdS_3$ is the group manifold of $SL(2,R)$; the action $S_L$ is the 
Wess-Zumino-Witten (WZW)
action for $SL(2,R)$ with level $Q_5$, 
and so possesses an affine $SL(2,R) \times SL(2,R)$ symmetry,
with independent generators for the left- and right-movers.
 
Euclidean $AdS_3$ is the coset manifold $SL(2,C)/SU(2)$; 
the action $S_E$ can be directly obtained from the $SL(2,C)$ WZW action
$S_{wzw}(g)$ \cite{gav}\footnote{This model has been studied in the past, 
owing to its relation to coset conformal field theories. It was 
shown in \cite{gk} that, when $G$ and $H$ are compact groups, 
the $G/H$ model is equivalent to the product of the $G$ model
and the $H^c/H$ model, where $H^c$ is the complexification of $H$,
when some BRST invariance is imposed on the product theory. It is
interesting to note that, if we take $G=H=SU(2)$, we find that
the topological $SU(2)/SU(2)$ model is equivalent to 
the product of the $SL(2,C)/SU(2) = AdS_3$ model and the
$SU(2) = S^3$ model (with the BRST invariance). Before imposing
the BRST invariance, the product model is nothing but the
worldsheet theory of strings on $AdS_3 \times S^3$. The
$SU(2)/SU(2)$ model may be useful to study some topological aspects
of the string theory in question.}.
The $SL(2,C)$ WZW model has two independent affine $SL(2,C)$ symmetries, 
associated with left- and right-movers. The quotient by
$SU(2)$ identifies the left and the right affine symmetries by
complex conjugation. This can be seen as follows.

We may regard the coset $SL(2,C)/SU(2)$ as the space of 
$2 \times 2$ hermitian complex matrices $h$ with unit determinant. 
To compare with the action (\ref{action}), we parametrize an $SL(2,C)/SU(2)$  matrix $h$ as
\beq
\label{sl2csu2}
    h = \left( \matrix{ e^{-\phi} + \gamma \bar{\gamma} e^{\phi}
          & e^{\phi} \gamma \cr
  e^\phi \bar{\gamma} & e^\phi \cr } \right).
\label{matrixh}
\eeq
The string action (\ref{action}) is simply the $SL(2,C)$
WZW action $S_{wzw}(h)$, with $h$ restricted to the form (\ref{sl2csu2}). By construction, the WZW action
$S_{wzw}(g)$ is invariant for arbitrary $g \in SL(2,C)$ under the left and the right $SL(2,C)$ symmetries
\beq
    g \rightarrow U(z) g V^\dagger(\bar{z}), ~~~U,V \in SL(2,C).
\eeq
However, $S_E = S_{wzw}(g=h)$ is
invariant only under the diagonal action 
\beq
   h \rightarrow U(z) h U^\dagger (\bar{z}), ~~~U \in SL(2,C),
\label{slaction}
\eeq
since $h$ is constrained to be hermitian.  The matrix $U$ is an arbitrary
holomorphic function of $z$; consequently, by Noether's theorem,
the corresponding currents $J^a$ ($a= \pm, 3$) are holomorphically 
conserved, 
\beq
\bar{\partial} J^a = 0.
\label{holomorphic}
\eeq 

So far we have discussed the classical symmetry of the action $S_E$.
The currents $J^a$ could receive quantum corrections,
but we expect that the conservation law (\ref{holomorphic}) still
holds. There are two instances in which quantum effects can be
perturbatively treated. 
 
\noindent
(A) When $Q_5$ is large, 
worldsheet quantum effects are suppressed
by $1/Q_5$.

\noindent
(B) If the functional integral 
is dominated by contributions at large $\phi$, we can use the
action
\beq
 S' = \frac{1}{4\pi} \int d^2 z
(\partial \phi \bar{\partial} \phi + \beta \bar{\partial} \gamma
   + \bar{\beta} \partial \bar{\gamma}
  - \beta \bar{\beta} e^{-2\phi/\alpha_+}
   - \frac{2}{\alpha_+} \phi \sqrt{g} R),
\label{largephi}
\eeq
where $\alpha_+ = \sqrt{2Q_5-4}$ and $R$ is the curvature of the worldsheet. 
The theory defined by the action $S'$ can be shown to be equivalent to the 
original one, upon integrating out $(\beta, \bar{\beta})$, taking into
account effects on the measure of the functional integral, and
rescaling the scalar fields by $\phi \rightarrow \phi \alpha_+$,
$\gamma \rightarrow \sqrt{2Q_5} \gamma$
 \cite{gav2}. 
For large $\phi$, the interaction term $\beta \bar{\beta} e^{-2\phi/\alpha_+}$
is suppressed and the free-field approximation to the fields
$(\beta, \gamma)$
becomes reliable. The $SL(2,C)$ currents in this notation are
given by
\beqa
   J^- &=&  \frac{1}{2} \beta \nonumber \\
   J^3 & = & \frac{1}{2}( \beta \gamma - \alpha_+ \partial \phi) \nonumber \\
   J^+ & = & \frac{1}{2} ( \beta \gamma^2 - 2 \alpha_+ \gamma \partial \phi
                             - \alpha_+^2 \partial \gamma). 
\label{currents}
\eeqa 
Moreover, because of the coupling of $\phi$ to the worldsheet
curvature $R$ in (\ref{largephi}), the effective string coupling
constant depends on the coordinate $\phi$ (the linear dilaton background).
For $\phi \rightarrow \infty$, the string coupling constant vanishes
asymptotically. Thus the spacetime
theory as well as the worldsheet theory is weakly coupled for
$\phi \rightarrow \infty$ in this picture \cite{gks}\footnote{
It may appear that, in the opposite limit $\phi \rightarrow -\infty$,
the effective string coupling constant diverges, 
and the spacetime theory is strongly coupled. This, however, is 
an artifact of the description in terms of the action $S'$.
In the limit $\phi \rightarrow -\infty$, the transformation 
relating $S$ to $S'$ breaks down, because the factor $e^{2\phi}$ 
multiplying the kinetic term for $\gamma$
in (\ref{action}) vanishes. In fact,
this transformation is an intermediate step
 in the T-duality transformation
along the isometry generated by a constant shift of 
$(\gamma, \bar{\gamma})$. 
(It becomes T-duality if we write $\beta = \partial
\tilde{\gamma}$ \cite{rove}.) The T-duality transformation is subtle when there
is a fixed point in the isometry. After T-duality, the dilaton
diverges at the fixed point, but this is an artifact, if the original
theory is well-defined at that point. This is the case for the string on 
$AdS_3$ since $\phi = -\infty$ is a regular boundary point on
$AdS_3$ and the string coupling is constant, $g_s^{-2} = Q_1\sqrt{Q_5}$ in the
original picture. In this paper, we will use $S'$ only when we analyze the
behavior of the functional integral for large $\phi$.}.

\subsection{Vertex Operators}

According to the ${\it AdS/CFT}$ duality, correlation functions
of ${\it CFT}$ correspond to string amplitudes on $AdS$ \cite{gkp,w}.
It is therefore useful to study vertex operators of the $AdS_3$ string.
Generally speaking, if a ${\it CFT}$ has a global affine $G$ symmetry, 
its vertex operators $V(z,\bar{z})$ 
take values in vector spaces representing the $G$ symmetry.
In the case of $AdS_3$, since the group $SL(2,C)$ is non-compact, we are led to consider infinite-dimensional representations as well as finite-dimensional ones. 
Teschner \cite{tes} introduced auxiliary coordinates $(x,\bar{x})$ to 
organize these representations.
Because $SL(2,C)$ acts on the matrix $h$ as $h \rightarrow U h U^\dagger$,
it is natural to consider the combination
\beq
  (1, -x) h \left(\matrix{1 \cr -\bar{x} \cr} \right)
   = e^{\phi/\alpha_+} (\gamma-x)(\bar{\gamma}- \bar{x})
    + e^{-\phi/\alpha_+},
\eeq
and to define the vertex operator $V_j$ 
by\footnote{The vertex operators of \cite{gks} correspond to the 
leading large $\phi$ part of the coefficients of the $x,\bar{x}$
expansion of $V_j$.}
\beq
   V_j(z,\bar{z}; x, \bar{x}) = \left( (\gamma - x) 
(\bar{\gamma} - \bar{x}) e^{\phi/\alpha_+} + 
e^{-\phi/\alpha_+} \right)^{2j}.
\label{vertex}
\eeq 
In the free-field approximation, it is straightforward to show that
this vertex operator gives the correct operator product expansion with
the $SL(2,C)$ currents,
\beq
   J^a(z) V_j(w,\bar{w}; x, \bar{x}) \sim
   \frac{1}{z-w} D^a  V_j(w,\bar{w}; x, \bar{x}),
\label{vertexope}
\eeq
where $a=3, \pm$, and
\beq
D^- = \frac{\partial}{\partial x},~~
D^3 = x \frac{\partial}{\partial x} 
                 - j,~~
D^+=x^2 \frac{\partial}{\partial x}
                 - 2jx.
\eeq
As we will show in section 6,  in evaluating the operator product
expansion of $J^a$ with $V_j$, we can
take $\phi$ to be arbitrarily large. 
Therefore the computation in (\ref{vertexope}) belongs to the 
case (B) discussed in section 3.1, and justifies the use
of the free-field approximation.

The global $SL(2,C)$ symmetry of \ads$_3$ corresponds to
the global conformal symmetry of the boundary \cft$_2$
generated by $L_0$ and $L_{\pm 1}$ \cite{mal,ms}. One can then relate
the highest weight $j$ of $SL(2,C)$ to the Virasoro highest
weight $h$ of the boundary \cft\ by 
\beq
      h = - j.
\eeq

In \cite{tes}, Teschner considered the case 
$j \in -1/2 + \sqrt{-1} {\bf R}$,
corresponding to principal representations of $SL(2,C)$.  
These are unitary representations and therefore appear in the
Hilbert space of the sigma model. 

In this paper, we are interested in the
situation when $h=-j$ is real, since $h$ is a conformal weight
of the boundary \cft$_2$. In this case,
the $SL(2,C)$ representation is non-unitary, and 
the corresponding supergravity mode is non-normalizable. 
For $h > 1/2$, because of the identity
\beq
\delta^{(2)}(z)=\frac{n-1}{\pi} \lim_{\epsilon\rightarrow 0} 
 \frac{\epsilon^{2n-2}}{ (\epsilon^2 + |z|^2)^n},
\eeq
the vertex operator $V_j$ behaves as
\beq
   V_{j=-h} \sim e^{2(h-1)\phi/\alpha_+} \delta^{(2)}(\gamma - x),
\label{bulkboundary}
\eeq
near the boundary $(\phi \rightarrow \infty)$ of $AdS_3$.
That is, the vertex operator $V_j$ has the same structure as 
the bulk-boundary Green's function used in the supergravity computation
of \cft\ correlation functions \cite{gkp,w}.  Of course,
this is not a coincidence. In the semi-classical approximation,
if a vertex operator $V_j$ is expressed as a function of 
$(\phi,\gamma,\bar{\gamma})$, the operator product
expansion (\ref{vertexope}) with
the $SL(2,C)$ currents implies that $V_j$ solves the supergravity wave equation
\be
(\Delta + j(j+1)) V_j = 0,
\label{wave}
\ee
where $\Delta$ is the Laplacian on \ads$_3$,
expressed in the coordinates $(\phi,\gamma,\bar{\gamma})$. 
The identification of vertex operators and bulk-boundary
Green's functions motivates us to interpret $(x,\bar{x})$ as
coordinates for the boundary \cft$_2$.

When $j$ is real, the vertex operator $V_j$ carries the $SL(2,C)$
weights $h=\bar{h}=-j$ and corresponds to a  
scalar field on \ads$_3$, such as a Kaluza-Klein excitation 
(on $S^3 \times M^4$)
of the dilaton field. To construct a vertex operator with $h \neq \bar{h}$, 
corresponding to tensor fields on $AdS_3$, we must 
include derivatives of the fields $(\phi, \gamma, \bar{\gamma})$.
Indeed,
we will see in section 5 that the graviton vertex operator corresponding to
the energy-momentum tensor $T(x)$ of \cft$_2$ is of this form.

We have found that the \ads/\cft\ duality implies that the vertex
operators $V_j$ belong to non-unitary representations 
of $SL(2,C)$, 
even though both the two-dimensional sigma model for Euclidean $AdS_3$
and the boundary \cft$_2$ are expected to be unitary theories, with 
Hilbert spaces of positive-definite inner product. Therefore
there is no state-operator correspondence in the sigma model\footnote{
A generalized version of the correspondence may hold if we
suitably extend the notion of states and allow for analytic
continuation of the quantum number $j$ in (3.9) \cite{tes,tes2}.}. 
This phenomenon is well-known in the Liouville model. In the
Liouville model, normalizable
states make up the Hilbert space and non-normalizable states
correspond to operators \cite{seib}.
In the $AdS_3$ case, unitary representations of $SL(2,C)$ are realized by
normalizable functions on $AdS_3$, whereas non-normalizable functions
give non-unitary representations. Thus it is in fact reasonable, by
the analogy with the Liouville model, that the vertex operators of the
sigma model belong to non-unitary representations.

\section{Semi-classical Analysis}

In this section\footnote{From now on we will work with
the original variables as they appear in (\ref{action}). Furthermore,
we will suppress the $Q_5$ dependence
until the discussion of
the central charge after equation (\ref{cc}).}
we will analyze correlation functions of the vertex operators
(\ref{vertex}) semi-classically. We propose the correspondence\footnote{This 
proposal is not complete as
it stands; see section~7 for a more precise statement.}
\be \label{corfie}
\langle\prod_i \int \, d^2 z_i V_{j_i}(z_i,\bar{z}_i;x_i,\bar{x}_i)\rangle_{\rm worldsheet} = 
\langle\prod_i V_{j_i}(x_i,\bar{x}_i) \rangle_{\rm boundary\,\,CFT} .
\ee
(In this expression, factors coming from the 
$S^3 \times M^4$ part of the target space are suppressed.)
Two facts directly support this proposal. First, there should be a 
one-to-one correspondence between vertex operators of the boundary \cft\ 
inserted at
specific boundary points and vertex operators of the worldsheet theory. Second,
according to (\ref{vertexope}), the worldsheet $SL(2,C)$ currents generate
the standard $SL(2,C)$ action on the boundary coordinates $x,\bar{x}$.

\begin{figure}[htb]
\begin{center}
\epsfxsize=3in\leavevmode\epsfbox{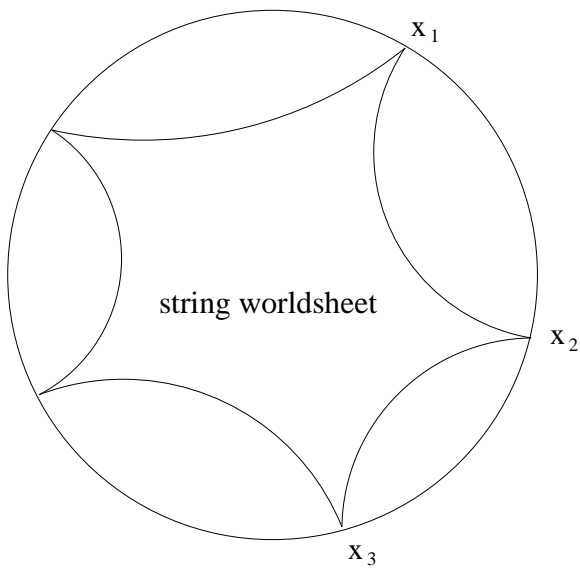}
\end{center}
\caption{Semi-classical worldsheet in the presence of vertex operators}
\label{F1}
\end{figure}

We obtain further insight in the structure of the correlation functions (\ref{corfie})
by studying the worldsheets that contribute to it in the semi-classical 
approximation.
The general solution to the equations of motion of (\ref{action}) in the absence of
sources is
\bea
\phi & = & \log(1+b(z) \bar{b}(\bar{z})) + \rho(z) + \bar{\rho}(\bar{z}) \nonumber \\
\gamma & = & a(z) + e^{-2 \rho(z)} \bar{b}(\bar{z})
(1+b(z) \bar{b}(\bar{z}))^{-1} 
 \nonumber \\
\bar{\gamma} & = & \bar{a}(\bar{z}) + e^{-2\bar{\rho}(\bar{z})} b(z) 
(1+b(z) \bar{b}(\bar{z}))^{-1},
\label{gensol}
\eea
for arbitrary holomorphic functions $a,b,\rho$. The case with sources can be dealt with
by allowing poles in $a,b,\rho$. To find these functions in the presence of 
arbitrary vertex operators is rather complicated (it is the analogue of the
uniformization problem in Liouville theory \cite{gimo}).
We will therefore only consider the behavior of the worldsheet near
a single vertex operator 
\be
V= \left( (\gamma-x)(\bar{\gamma}-\bar{x}) e^{\phi} + e^{-\phi} \right)^{2j}
(z_0)
\ee
at the point $z=z_0$. The relevant equations of motion read
\bea
\frac{1}{2\pi} \partial \bar{\partial} \phi - 
\frac{1}{2\pi} e^{2\phi} 
\partial \bar{\gamma} \bar{\partial} \gamma + 
2 j \frac{(\gamma-x)(\bar{\gamma}-\bar{x}) e^{\phi} -
 e^{-\phi}}{ (\gamma-x)(\bar{\gamma}-\bar{x}) e^{\phi} + e^{-\phi} }
 \delta^{(2)}(z-z_0) & = & 0 \\
\frac{1}{4\pi} \partial (e^{2\phi} \bar{\partial} \gamma ) 
+ 
2 j \frac{(\gamma-x)e^{\phi} 
 }{ (\gamma-x)(\bar{\gamma}-\bar{x}) e^{\phi} + e^{-\phi} }
 \delta^{(2)}(z-z_0) & = & 0 \\
\frac{1}{4\pi} \bar{\partial}(e^{2\phi} \partial \bar{\gamma} ) 
+ 
2 j \frac{(\bar{\gamma}-\bar{x}) e^{\phi} 
 }{ (\gamma-x)(\bar{\gamma}-\bar{x}) e^{\phi} + e^{-\phi} }
 \delta^{(2)}(z-z_0) & = & 0.
\eea
This system has the solution
\bea
\phi & = & 2 j \log|z-z_0|^2 + b + c(z-z_0) + \bar{c} (\bar{z}-
\bar{z}_0) + \ldots  \nonumber \\
\gamma & = & x + a(z-z_0)^{-4j}(\bar{z}-\bar{z}_0)^{1-4j} 
- 2 a c (z-z_0)^{1-4j}(\bar{z}-\bar{z}_0)^{1-4j} + \ldots \nonumber \\
\bar{\gamma} & = & \bar{x} + \bar{a} (z-z_0)^{1-4j}(\bar{z}-\bar{z}_0)^{-4j}
-2 \bar{a} \bar{c} (z-z_0)^{1-4j}(\bar{z}-\bar{z}_0)^{1-4j} + \ldots,
\label{semsol} 
\eea
where $a,b,c$ are some arbitrary constants and the dots indicate higher-order
regular terms. The corresponding functions in (\ref{gensol})
are
\be a(z)=x, \qquad b(z)=a e^b (z-z_0)^{1-4j}, \qquad \rho(z)=2 j \log(z-z_0) + \frac{b}{2} + c(z-z_0).  
\ee
Since we consider only vertex operators with $j\leq -1/2$, corresponding to boundary
conformal weight $h\geq 1/2$, the worldsheet coordinates at $z_0$ are 
$(\phi,\gamma,\bar{\gamma})(z_0)=(\infty,x,\bar{x})$. Thus the worldsheet
develops an infinite tube that 
attaches to the point $(x,\bar{x})$ at the boundary
of \ads$_3$. In the field theory limit, the worldsheet degenerates, and 
we recover the picture of \cite{w}, where boundary correlation functions
are expressed in terms of Feynman diagrams consisting of bulk-bulk and
bulk-boundary propagators.  
This is further evidence for the identification
(\ref{corfie}). The structure of the worldsheet is illustrated in 
figure \ref{F1}.

When we evaluate the semi-classical contribution to the correlation 
function (\ref{corfie}),
we encounter a divergence arising from the stretching of the worldsheet
to the boundary at infinity of \ads$_3$. 
To regularize this divergence,
we introduce a worldsheet UV cutoff $\epsilon$, and multiply the correlation function
by a suitable power of $\epsilon$ before taking the limit $\epsilon\rightarrow 0$. The appropriate power
is easily determined (see \cite{seib} for a similar analysis for Liouville theory) and 
corresponds to a wave function renormalization for each vertex operator $V_j$,
\be V_j^{ren} = {\epsilon}^{8 j^2} V_j.
\label{regularization}
\ee
A similar renormalization has also been found in \cite{gav2}, where correlation functions
of $V_j$ with $j>0$ were studied. In that situation, one consequence of the renormalization was that the
$e^{-\phi}$ in $V_j$ could be dropped, leading to an exact free-field 
representation of the correlation functions. It should be possible to find similar exact free-field representations of the
correlation functions of $V_j$ with $j<0$, because $SL(2,C)$ representations
with spins $j$ and $-1-j$ are equivalent. We shall not pursue this further here; nevertheless, we will find that the wave
function renormalization brings about many simplifications. In particular, it explains
why the free-field approximation is valid, and plays a crucial role in proving
the Virasoro Ward identities of the boundary \cft.

\begin{figure}[htb]
\begin{center}
\epsfxsize=3in\leavevmode\epsfbox{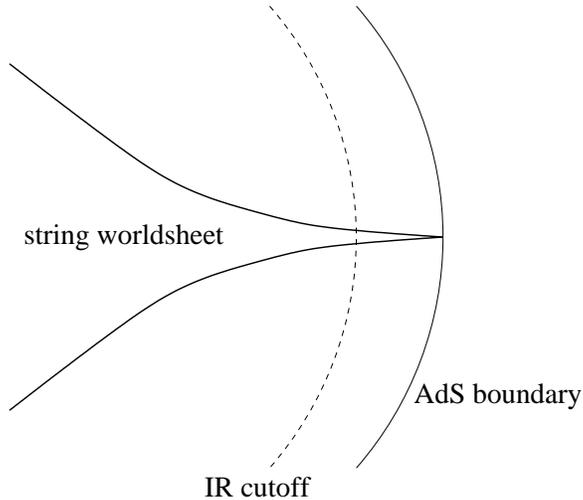}
\end{center}
\caption{Bulk IR cutoff versus worldsheet UV cutoff}
\label{F2}
\end{figure}

Besides $\epsilon$, there are two other cutoffs in the problem, 
the IR cutoff of the bulk theory and the UV cutoff of the boundary
\cft. All three cutoffs are related.
According to (\ref{semsol}), the bulk IR cutoff $U_0$ 
in $U=e^{\phi}$ is 
\beq
U_0=\epsilon^{4j},
\eeq
 and depends on which
vertex operator is inserted. The UV cutoff $\tilde{\epsilon}$ of
the boundary \cft\ is 
related to $U_0$ by \cite{iruv}
\beq
\tilde{\epsilon}=U_0^{-1} .
\eeq
With this identification of the cutoff parameters, (\ref{regularization})
may be expressed in terms of the boundary \cft\ cutoff $\tilde{\epsilon}$ as
\beq
   V_j^{ren} = \tilde{\epsilon}^{2h} V_j.
\eeq
The factor $\tilde{\epsilon}^{2h}$ matches the scaling
behavior of the primary field of the boundary \cft\ corresponding
to the worldsheet vertex operator $V_j$.
This fits well with the \ads/\cft\ duality\footnote{
These relations among the cutoff parameters hold even when we
restore the $Q_5$ dependence.}.   
The relation between $U_0$ and $\epsilon$ is illustrated in 
figure~\ref{F2}.

We next turn to the fluctuations around the semi-classical worldsheet. 
If we denote
the semi-classical worldsheet by $(\phi_0(z),\gamma_0(z),\bar{\gamma}_0(z))$
and quantum fluctuations by $(\phi_q(z),\gamma_q(z),\bar{\gamma}_q(z))$, 
we see from (\ref{semsol}) that the dominant contribution to
the kinetic term of the quantum fields near $z=z_0$ is
%
\be
\int d^2 z |z-z_0|^{-2} (\phi_q(z)^2 + |z-z_0|^{8j} |\gamma_q(z)|^2).
\ee
For the action to be finite, we need 
\be 
\label{semi2}
\phi_q(z_0) \sim \epsilon^{\frac{1}{2}}, \qquad 
\gamma_q(z_0) \sim \epsilon^{\frac{1}{2}-4j} .
\ee
In particular, the fluctuations of the worldsheet vanish near the boundary
as we take $\epsilon\rightarrow 0$. Furthermore, no quantum terms in 
the background
field expansion of the vertex operators $V_j$ contribute to the correlation
function (\ref{corfie}). Thus the one-loop worldsheet correction to the correlation
function consists only of the determinant of the kinetic term of the quantum fields
$(\phi_q(z),\gamma_q(z),\bar{\gamma}_q(z))$.

\section{The Virasoro Algebra}

So far we have discussed the primary fields of the boundary \cft. 
We now turn our
attention to the boundary Virasoro algebra. Let us briefly recall 
how the Virasoro
algebra arises in \cite{bh}. First, we define spaces that are asymptotically anti-de Sitter
by imposing on the metric the boundary conditions 
\be G_{\phi\phi} = 1 + {\cal O}(e^{-2\phi}),  \qquad
G_{\phi\gamma}=G_{\phi\bar{\gamma}} = {\cal O}( e^{-2\phi}) \ee
\be G_{\gamma\gamma}=G_{\bar{\gamma}\bar{\gamma}} = {\cal O}(1), \qquad 
 G_{\gamma\bar{\gamma}}=\frac{1}{2} e^{2\phi} + {\cal O}(1). \ee
Next, we consider the group $G$ of diffeomorphisms that preserve these boundary conditions. 
To each of these one can associate an ADM-type charge that vanishes identically 
for a subgroup $H$ of diffeomorphisms that decay sufficiently fast at
infinity. The algebra of the quotient $G/H$ is the Virasoro algebra.
The infinitesimal diffeomorphisms corresponding to the generators $L_n$ are
\bea
\xi^{\gamma} & = & -\gamma^{n+1}+ {\cal O}( e^{-4\phi})  \nonumber \\
\xi^{\bar{\gamma}} & = & \frac{1}{2} n (n+1) \gamma^{n-1} e^{-2\phi}+
{\cal O}( e^{-4\phi})  \nonumber \\
\xi^{\phi} & = & \frac{1}{2} (n+1) \gamma^n + {\cal O}( e^{-2\phi}) .
\label{virgen}
\eea
We have given only the holomorphic part of the Virasoro algebra---the full Virasoro algebra consists
of the sum of these generators and their complex conjugates.  Moreover, our choice of generators is not unique---we could equally well replace $\gamma$ by $\gamma - \gamma_0$ everywhere.

If we perform one of the infinitesimal diffeomorphisms (\ref{virgen}) in the worldsheet
theory, the result is the insertion of a combined vertex
operator for the graviton and the {\it NS}-{\it NS} two-form field.  
This vertex operator is given by
\be
L_n = \delta S_n=
\int d^2 z \left(\frac{1}{2}(n+1)n 
\gamma^{n-1} (\partial\gamma \bar{\partial} \phi -
 \bar{\partial} \gamma \partial \phi) + \frac{1}{2} (n+1)n(n-1) \gamma^{n-2} 
\partial \gamma \bar{\partial} \gamma \right).
\ee
We have neglected subleading terms in (\ref{virgen}).  

Normally, the graviton vertex operator corresponding 
to a diffeomorphism is on-shell
BRST exact and decouples from the theory, as it corresponds to an unphysical graviton.
Alternatively, 
the graviton vertex operator is the sum of a total derivative and
equation of motion terms, and the latter can be dropped by the canceled
propagator argument \cite{pol}. 

In the case of \ads$_3$, however, something special happens. 
Although we can formally write $\delta S_n$ as $\{Q_{BRST},X\}$, $X$ is 
not normalizable, and therefore $\delta S_n$ is a non-trivial element of the BRST cohomology. 
Alternatively, as we will show below, the total derivative
terms cannot be dropped: in fact, these terms give rise to the contour integral representation
of the Virasoro generators of \cite{gks}. 
From either perspective, then, the vertex operators $\delta S_n$ 
are physical states of the theory.  Since there are no propagating 
gravitons in three
dimensions, they correspond to degrees of freedom living purely 
on the boundary of
\ads$_3$ ($i.e.$, singletons). 

Altogether we are led to identify an insertion of 
the boundary stress-energy tensor $T(x)$ in a boundary
correlation function with the insertion of the vertex operator
$T(\phi,\gamma,\bar{\gamma}; x)$ in the
worldsheet correlation function given by
\bea
T(x)&=& \sum_{n=-2}^{-\infty} L_n x^{-n-2} \nonumber \\
& =& 
\int d^2 z \left(\frac{1}{(\gamma-x)^3} (\partial\gamma \bar{\partial} \phi -
 \bar{\partial} \gamma \partial \phi) - \frac{3}{(\gamma-x)^4} 
\partial \gamma \bar{\partial} \gamma \right).
\label{defT}
\eea

We saw previously in (\ref{bulkboundary}) that, for large
$\phi$, vertex operators behave like
bulk-boundary Green's functions, and in particular that they
become localized at single points. The same is true for
the stress-energy tensor, although this is less obvious from
 (\ref{defT}). Consider for definiteness the
second term in (\ref{defT}). For large $\phi$, this term
seems to be subleading compared to the term $e^{2\phi}
\partial \bar{\gamma} \bar{\partial} \gamma$. However, we
must be careful, because $(\gamma-x)^{-4}$ blows up near
$\gamma=x$. Up to terms subleading in $e^{-2\phi}$, the 
second term in (\ref{defT}) can be rewritten as\footnote{
For example, one can choose the representative
$$
\xi^{\bar{\gamma}} = \frac{\bar{\gamma}-\bar{x}}{(\gamma - x)^2}
\frac{e^{-2\phi}}{|\gamma-x|^2 + e^{-2\phi}} 
$$      
in place of (6.1) to find the expression (5.6) for large $\phi$.} 
\be \label{aux1}
-3 \int d^2 z e^{2\phi} \left( 
\frac{(\bar{\gamma}-\bar{x})^2}{(\gamma-x)^2} 
\frac{e^{-2\phi}}{(|\gamma-x|^2 + e^{-2\phi})^2} \right)
\partial \gamma \bar{\partial} \gamma .
\ee
Since the Brown-Henneaux diffeomorphisms are defined up to
subleading terms only, the same is true for $T$, and we might
as well have used (\ref{aux1}) in our definition of $T$. 
For large $\phi$, (\ref{aux1}) behaves as
\be
-3 \int d^2 z e^{2\phi}  \left(
\frac{(\bar{\gamma}-\bar{x})^2}{(\gamma-x)^2} 
\delta^{(2)}(\gamma-x) \right) 
\partial \gamma \bar{\partial} \gamma .
\label{bulkboundary2}
\ee
This is the analogue of (\ref{bulkboundary}) for the
stress tensor. As in (\ref{bulkboundary}), it behaves like
a bulk-boundary Green's function, and is localized
on the boundary of $AdS_3$.

\section{Boundary Ward Identity}

As a first application of the definition (\ref{defT}), we will show that 
it correctly
reproduces the Virasoro Ward identities of 
the boundary \cft. We first discuss the case 
of a single insertion of the stress-energy tensor and an 
arbitrary number of primary 
fields. The case with more than one stress tensor insertion is more complicated and
will be discussed later. 

Our strategy for proving the Virasoro Ward 
identities is to perform a change of variables
in the path integral corresponding to a Brown-Henneaux diffeomorphism. 
The diffeomorphism
corresponding to $T(x)$ is
\bea
\xi^{\gamma} & = & -\frac{1}{\gamma-x} + {\cal O}( e^{-4\phi})  \nonumber \\
\xi^{\bar{\gamma}} & = & \frac{1}{(\gamma-x)^3} e^{-2\phi}+
{\cal O}( e^{-4\phi})  \nonumber \\
\xi^{\phi} & = & \frac{-1}{2(\gamma-x)^2} + {\cal O}( e^{-2\phi}) .
\label{virgen2}
\eea    
Let us perform this change of variables on the correlation function
\be
\langle\prod_i \int \, d^2 z_i V_{j_i}(z_i,\bar{z}_i;x_i,\bar{x}_i)
\rangle_{\rm worldsheet} . 
\ee
There are two contributions: one comes from the variation of the action,
yielding $T(\phi,\gamma,\bar{\gamma};x)$, while the other comes from the
variation of the vertex operators and has the form
\bea
\delta_{\xi} V_{j_i} & = & -\left( 
\frac{-j_i}{(x-x_i)^2} + \frac{1}{(x-x_i)} \frac{
\partial}{\partial x_i} \right) V_{j_i}(x_i) \nonumber \\
& & - \frac{j_i (\gamma-x_i)^2}{(\gamma-x)^3 (x-x_i)^2 } 
 (e^{\phi}(\gamma-x_i)(\bar{\gamma}-
\bar{x}_i) + e^{-\phi})^{2j_i-1} R,
\label{varv}
\eea
where
\be
R=e^{-\phi} (\gamma-3 x + 2 x_i) 
+ e^{\phi} (\gamma-x_i) (\bar{\gamma}-\bar{x}_i) (\gamma-x).
\ee
In the first line of (\ref{varv}) we recognize the operator 
product expansion of $T(x)$ with $V_{j_i}(x_i)$. Using the results
(\ref{semsol}) and
(\ref{semi2}) from the semi-classical analysis, we determine that the
remainder, $i.e.$, the second line in (\ref{varv}), gives a vanishing
contribution to the correlation function. Indeed, the leading term in
the background field expansion vanishes, as do all terms containing
quantum fields, after taking into account the renormalization factor
(\ref{regularization}). The main reason for this is the explicit factor of
$(\gamma-x_i)^2$ in the second line of (\ref{varv}). 

We have shown that 
\be
\langle T(\phi,\gamma,\bar{\gamma};x) \prod_i \int \, d^2 z_i 
V_{j_i}(z_i,\bar{z}_i;x_i,\bar{x}_i)\rangle_{\rm worldsheet}
\ee
is equal to 
\be
\sum_i \left(\frac{h_i}{(x-x_i)^2} + \frac{1}{(x-x_i)} 
\frac{\partial}{\partial x_i} \right)
\langle \prod_i \int \, d^2 z_i V_{j_i}(z_i,\bar{z}_i;x_i,\bar{x}_i)\rangle_{\rm worldsheet},
\ee
where $h_i = -j_i$. 
Since both correlation functions have a corresponding meaning
in the boundary \cft, this proves the Virasoro Ward identities of
the boundary \cft, to all orders in the string worldsheet theory.

This analysis confirms that only
the leading large $\phi$ behavior of the Brown-Henneaux diffeomorphisms
is relevant. Had we chosen any other representative, we
would still have obtained the correct Virasoro Ward identity.
This is because the insertion of a graviton vertex operator corresponding 
to a diffeomorphism that decays faster, at large
$\phi$, than the Brown-Henneaux diffeomorphism automatically yields
zero. Again, this is as expected.

We can now also make contact with the contour representation of the
Virasoro generators in \cite{gks}. To do this, we rewrite $T$ in
(\ref{defT}) as the sum of total derivative and equation of
motion terms. The equation of motion terms can be dropped 
if we view the UV regularization as cutting 
discs of radius $\epsilon$ out of the worldsheet
around each of the vertex 
operators $V_{j_i}$. The equation of motion terms have only
contact-term interactions with the $V_{j_i}$, and can
therefore be neglected. What remains is the total derivative
terms. In the presence of
the $V_{j_i}$, the regularized worldsheet acquires a boundary,
consisting of the boundaries of the small discs.
The total derivative terms thus turn into a sum of contour
integrals encircling each of the
vertex operators. The relevant contour integrals
for $T(x)$ are
\be \label{contour}
 \sum_i \oint_{z_i} dz
\left(\frac{-1}{\gamma-x} e^{2\phi} \partial \bar{\gamma} +
\frac{-1}{2(\gamma-x)^2} \partial \phi \right) 
+ \oint_{z_i} d\bar{z} \left(\frac{-1}{(\gamma-x)^3} \bar{\partial} \gamma +
  \frac{1}{2(\gamma-x)^2} \bar{\partial}\phi\right) 
\ee
These contour integrals are just the canonical worldsheet generators
of the Brown-Henneaux diffeomorphisms. Therefore, the contour
integral can be worked out semi-classically, resulting in
in (\ref{varv}). All corrections to this semi-classical
result vanish as we take the regulator to zero. 
The contour integral representation of the Virasoro generators
in \cite{gks} is a slight modification of (\ref{contour}),
namely,
\be \label{contour2}
 \sum_i \oint_{z_i} dz
\left(\frac{-1}{\gamma-x} e^{2\phi} \partial \bar{\gamma} 
+\frac{-1}{(\gamma-x)^2} \partial \phi
 + \frac{1}{(\gamma-x)^3} {\partial} \gamma \right).
\ee   
The difference between (6.7) and (6.8) is annihilated
when acting on $V_{j_i}$'s.
In the free-field approximation, the integrand of (6.8)
contains purely holomorphic operators, and it is valid to use
free-field OPE's in computing contour integrals
around the $V_{j_i}$. Again we recover (\ref{varv}) up to terms that
vanish as the regulator is taken to zero. This shows precisely
how and when the free-field representation is exact.

\section{$T(x)T(y)$ OPE and Central Charge}

To evaluate the insertion of two or more boundary stress tensors in
a correlation function, one might consider, along the
lines of the above procedure, 
performing consecutive Brown-Henneaux diffeomorphisms
and studying the resulting Ward identities. 
The only novel feature would be the variation
of the stress tensor under a Brown-Henneaux diffeomorphism.
As it will turn out, 
this is not the whole story and has to be supplemented by an additional
ingredient. 
The variation of the stress tensor can be computed using 
the contour integral representation
(\ref{contour}). It is easiest to vary a mode of
(\ref{contour}),
\bea
L_n &\equiv& \sum_i \oint_{x_i} dz
\left(-\gamma^{n+1} e^{2\phi} \partial \bar{\gamma} 
+\frac{1}{2}(n+1)\gamma^n \partial \phi\right) 
+ \nonumber \\
&& + \sum_i \oint_{x_i} d\bar{z} \left(-\frac{1}{2}n(n+1)\gamma^{n-1} 
\bar{\partial} \gamma -
  \frac{1}{2}(n+1)\gamma^{n} \bar{\partial}\phi \right) ,
\eea
under the Brown-Henneaux diffeomorphism
(\ref{virgen}) corresponding to $L_m$. This yields
\bea
\delta_m L_n & = & (m-n)L_{m+n} - 
(m^3-m)\sum_i \oint_{z_i} \gamma^{m+n-1} \partial\gamma +
\nonumber \\
& & +\frac{1}{2} m(m+1) \sum_i \left(\oint_{z_i} dz 
\gamma^{n+m} \partial \phi +
\oint_{z_i} d\bar{z} \gamma^{n+m} \bar{\partial} \phi\right)+ \nonumber \\
& & + \frac{1}{4} m(m+1)(n + 2m -1) \sum_i 
\left(\oint_{z_i} dz \gamma^{n+m-1} \partial \gamma +
\oint_{z_i} d\bar{z} \gamma^{n+m-1} \bar{\partial} \gamma\right) .
\label{virvar}
\eea
The two last lines in this expression vanish as
we send the regulator to zero. The last term in the first line
is similar to the expression for the central charge proposed in
\cite{gks}. However, since we insert the boundary Virasoro generators
at points different from the insertion points of the primary fields,
this term does not contribute. The $L_n$ correspond to insertions
of $T$ at $0$ or $\infty$, and 
\beq
\oint_{z_i} \gamma^{m+n-1} \partial\gamma=0,
\eeq
if $x_i\neq 0,\infty$. 

\begin{figure}[htb]
\begin{center}
\epsfxsize=3in\leavevmode\epsfbox{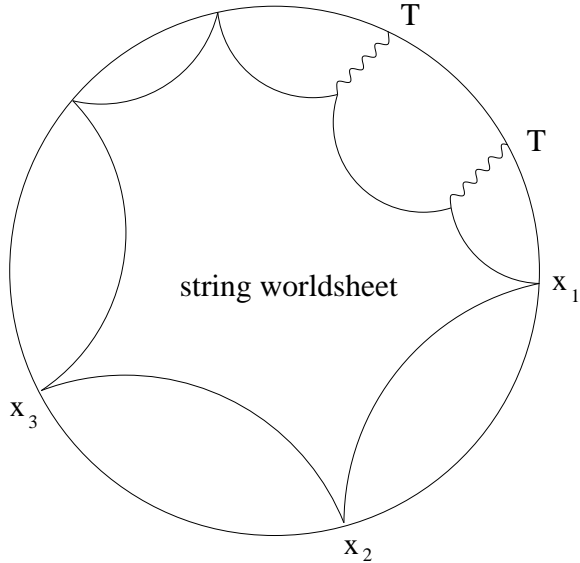}
\end{center}
\caption{A single string worldsheet contributing
to the $\langle TTV_1 \ldots V_n\rangle$ correlator. 
This diagram does not contribute the
central charge of the Virasoro algebra.}
\label{F3}
\end{figure}

All that remains from (\ref{virvar}) is the Virasoro algebra with zero
central charge. Therefore, performing two Brown-Henneaux variations
gives us the correct Ward identity for the insertion of two
stress tensors in a correlation function of primary fields, except
for the central charge term. 

\begin{figure}[htb]
\begin{center}
\epsfxsize=3in\leavevmode\epsfbox{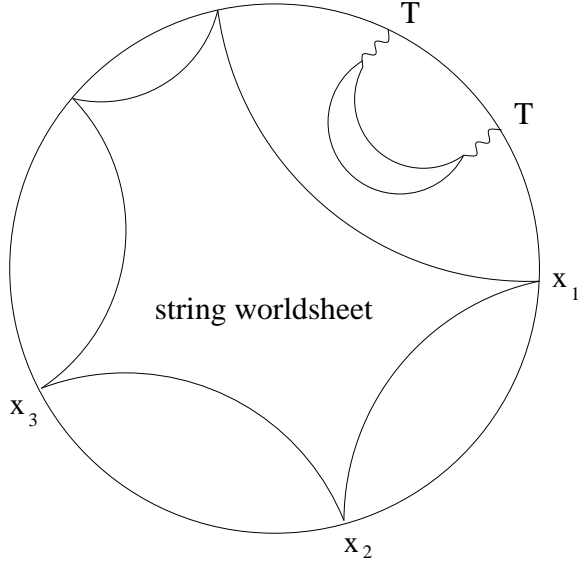}
\end{center}
\caption{A multiple string worldsheet contributing
to the $\langle \langle TTV_1 \ldots V_n\rangle\rangle$ correlator.
The central charge $c=6Q_1Q_5$ is obtained from this diagram.}
\label{F4}
\end{figure} 

The reason that the computation does not capture  
the central charge in this Ward identity
is the following. In the \ads/\cft\ duality, the string
theory on \ads\ is second-quantized. Therefore we need to
 sum over all possible string worldsheets, including disconnected
ones. This corresponds in the supergravity limit \cite{w}
to the prescription to sum over all Feynman
diagrams constructed out of bulk-bulk and bulk-boundary
propagators, including disconnected Feynman diagrams. So far we
have been focusing on a single string worldsheet, as 
illustrated in figure~\ref{F3}. Let us denote
by $\langle \langle V_1 \ldots V_n\rangle\rangle$ the second-quantized
string theory correlation function
involving arbitrary multiple worldsheets, and by 
$\langle V_1 \ldots V_n\rangle$ the correlation function obtained from
a single worldsheet. Then
\bea \langle \langle V_1 \ldots V_n\rangle\rangle & = & 
\langle V_1 V_2 \cdots V_n \rangle +
\langle V_1 V_2 \rangle \langle V_3 \cdots V_n\rangle + \cdots.
\eea
It is $\langle \langle V_1 \ldots V_n\rangle\rangle$, rather
than $\langle V_1 
\ldots V_n\rangle$, that should 
be identified with a boundary \cft\
correlation function. One can easily check that
the Virasoro Ward identities still hold if 
we replace $\langle V_1 \ldots V_n\rangle$
by $\langle \langle V_1 \ldots V_n \rangle \rangle$.
 However, the correlation function  $\langle \langle  TT V_1 
\ldots V_n\rangle\rangle$ containing two boundary stress-energy
tensors includes a contribution from
\be \langle TT\rangle\langle \langle V_1 \ldots V_n\rangle\rangle, \ee
as illustrated in 
figure~\ref{F4}. We have not yet computed 
the two-point function of stress tensors.
The previous analysis of the Ward identities
does not apply to $\langle TT\rangle$, because the contour 
integral representation
of $T$ cannot be used in the absence of other vertex operators. 

When $Q_5 \gg 1$,  the two-point function of
the energy-momentum tensor is computable in the semi-classical
approximation giving
\beq
  \langle T(x) T(y) \rangle_{{\rm worldsheet}} =
  \frac{c/2}{(x-y)^4},
\label{cc}
\eeq
with $c = 6Q_1Q_5$. Let us outline the derivation of this formula. 
As shown in
section 5, the energy-momentum tensor $T(\phi, \gamma, \bar{\gamma};
x)$ can be interpreted as the bulk-boundary Green's function
for a graviton in $AdS_3$. Therefore, in the semi-classical approximation,  
$\langle T(x) T(y) \rangle_{{\rm worldsheet}}$   
can be identified with the two-point graviton amplitude in 
the $AdS_3$ supergravity. 
The relevant part of the supergravity action is (up to numerical
coefficients)
\beq
  S = \frac{1}{l_p} \int  d\phi d\gamma d\bar{\gamma}
                   \sqrt{g} (R + l_{AdS}^{-2}) 
         + ({\rm boundary~term}).
\eeq
If we perturb the metric by $g_{\mu\nu} \rightarrow g_{\mu\nu} 
+ h_{\mu\nu}$, the action is expanded as
\beq
  S = \frac{1}{l_p} \int d\phi d\gamma d\bar{\gamma}
                   \sqrt{g} \partial h \partial h + \cdots.\
\eeq
Let us choose $h$ to be the bulk-boundary Green's function 
with sources at $x$ and $y$ on the boundary.
Since $\sqrt{g} \sim l_{AdS}^3$ and $\partial^2 \sim l_{AdS}^{-2}$,
the action  scales as $S \propto l_{AdS}/l_p$.
The $x, y$ dependence of the action is determined by the
$SL(2,C)$ invariance, and we obtain\footnote{An explicit
computation of this can be found in \cite{hkl}.}
\beq
  S \sim \frac{l_{AdS}/l_p}{(x-y)^4} \sim
          \frac{Q_1 Q_5}{(x-y)^4}.
\eeq

Thus the Virasoro central charge indeed arises from 
the two-point graviton amplitude, which is
a part of the disconnected diagram in figure 4. 

It should also be possible to obtain (\ref{cc}) directly
from a string worldsheet computation. 
In string theory, every genus-zero worldsheet 
carries an extra factor of 
$g_s^{-2}$. Therefore the disconnected diagram of figure 4 has
an extra factor of $g_s^{-2} = Q_1 \sqrt{Q_5}$ compared to the
connected diagrams of figure 3. The worldsheet amplitude itself
is a function of $l_{AdS}/l_s = \sqrt{Q_5}$ only. For the
two-point function of the energy-momentum tensors, our
preliminary computation (analogous to the spacetime computation
in \cite{hesk}) indicates that the only $l_{AdS}$ dependence comes
from the measure of the $\phi$ zero mode integral. Thus
we expect that this computation also reproduces (\ref{cc})
with $c \sim Q_1 Q_5$. It would be desirable to make this 
computation more precise in order to estimate
finite $Q_5$ corrections to the central charge formula.

\section{Discussion}

In this paper we have studied string theory on \ads$_3$ and found that
many properties of the \ads/\cft\ duality can be
understood from a semi-classical analysis. In particular,
we found vertex operators in the worldsheet theory that correspond
to the insertion of operators in the boundary \cft. The structure
of these vertex operators is somewhat reminiscent of the 
master field for large $N$ field theory. We showed that the 
string worldsheet stretches
to the boundary of \ads$_3$ in the presence of such vertex 
operators, and that the Virasoro generators of Brown and Henneaux
directly give rise to the contour integral representation
of the Virasoro algebra in \cite{gks}. We have explained why,
in this representation, the contour is localized
near the boundary of \ads$_3$, and deduced from this
the Virasoro Ward identities of the boundary theory. 
This clarifies several aspects of \cite{gks}. However, in our
formulation the central charge arises by a different
mechanism than one put forth in \cite{gks}. We found no need to
introduce fundamental strings at infinity and 
to consider worldsheets wrapping a certain number of times
around the boundary of \ads. Instead the central charge arose
from the disconnected diagram of the second-quantized string theory.
It is conceivable that the
two different pictures of the central charge are roughly
analogous to the short and long string pictures that one encounters,
for instance, in matrix string theory \cite{motl,bs,dvv}. 
The precise meaning and
definition of such a long string picture would require
further clarification.

Several other issues deserve further investigation. We have not yet given
a detailed derivation of the central charge from the worldsheet
theory. It would be interesting to do this and to see whether
the central charge satisfies a non-renormalization theorem
in the case of superstrings on $AdS_3 \times S^3 \times M^4$.
In addition, we would like to extend this analysis to Lorentzian
signature \ads$_3$, and to have a more detailed understanding of
the spectrum and the vertex operators in that case. 
Finally, we would like to see whether this formulation
of string theory on \ads$_3$ can be used in a practical
way to compute higher order $\alpha'$ corrections to 
supergravity results.

\bigskip

\noindent
{\bf Note Added:}

Toward the completion of this paper, we received 
\cite{kll}. In that paper, string theory
on \ads$_3 \times S^3 \times T^4$ is studied using the approach of
\cite{gks}, and a disagreement is found in the spectra
of the $U(1)^4$ charges between the string theory on 
\ads$_3$ and the \cft$_2$ with target space $(T^4)^N/{\cal S}_N$. 
Since that computation
depends crucially on the evaluation of the $U(1)$ central
charge, it would be interesting to calculate the central
charge from our point of view and see if the disagreement
persists. 

\section*{Acknowledgments}

We would like to thank K. Bardak\c{c}i for discussions
and J. Teschner for useful correspondence.
JdB would like to thank S. Shatashivili for collaboration
at an earlier stage of this work. 
This work was supported in part by the NSF
grant PHY-95-14797 and the DOE grant DE-AC03-76SF00098.

\end{document}